\documentclass[preprint,aps,10pt]{revtex4}
\pdfoutput=1
\usepackage{latexsym}
\usepackage{amsmath}
\usepackage{times}
\usepackage{amssymb}
\usepackage{fancyheadings}
\usepackage[T1]{fontenc}
\usepackage[utf8]{inputenc}
\usepackage{url}
\usepackage{hyperref}
\usepackage{color}
\usepackage{graphicx}
 \usepackage{epsfig}
\usepackage{mathptmx}
\usepackage{bm}
\usepackage{array}
\usepackage{algorithm}
\usepackage{algorithmic}
\usepackage{url}
\usepackage{hyperref}
\usepackage{algorithm}
\usepackage{algorithmic}
\usepackage{color}

\usepackage{lscape}
\usepackage[normalem]{ulem}

\usepackage{natbib}
\begin{document}
\title{The critical Barkhausen avalanches in thin random-field 
  ferromagnets with an open boundary}

%\author{Svetislav Mijatovi\'c$^1$, Sanja Jani\'cevi\'c$^1$, Djordje
%  Spasojevi\'c$^1$, Geoff J. Rodgers$^2$, Bosiljka Tadi\'c$^{3,4}$}
%\affiliation{ %\flushleft
%$^1$Faculty of Physics; University of Belgrade; Belgrade; Serbia;
%$^2$Brunel University  West London UK;
%$^3$Department  for Theoretical Physics; Jo\v{z}ef Stefan Institute;
%P.O. Box 3000; SI-1001 Ljubljana; Slovenia and Complexity Science Hub;
%Vienna; Austria\\ \hspace{1cm}}

\author{Bosiljka Tadi\'c$^{1}$, Svetislav Mijatovi\'c$^2$, Sanja Jani\'cevi\'c$^2$, Djordje
  Spasojevi\'c$^2$, Geoff J. Rodgers$^3$}
\affiliation{ %\flushleft
$^1$Department  for Theoretical Physics; Jo\v{z}ef Stefan Institute;
P.O. Box 3000; SI-1001 Ljubljana; Slovenia and Complexity Science Hub;
Vienna; Austria;
$^2$Faculty of Physics; University of Belgrade;  POB 368, 11001 Belgrade; Serbia;
$^3$Brunel University London, Uxbridge Middlesex UB8 3PH UK\\ 
\hspace{1cm}}

\begin{abstract}
The interplay between the critical fluctuations and the sample geometry is investigated numerically using thin random-field ferromagnets exhibiting the field-driven magnetisation reversal on the hysteresis loop. The system is studied along the theoretical critical line in the plane of random-field disorder and thickness. The thickness is varied to consider samples of various geometry between a two-dimensional plane and a complete three-dimensional lattice with an open boundary in the direction of the growing thickness. We perform a multi-fractal analysis of the Barkhausen noise signals and scaling of the critical avalanches of the domain wall motion. Our results reveal that, for sufficiently small thickness, the sample geometry profoundly affects the dynamics by modifying the spectral segments that represent small fluctuations and promoting the time-scale dependent multi-fractality. Meanwhile, the avalanche distributions display two distinct power-law regions, in contrast to those in the two-dimensional limit, and the average avalanche shapes are asymmetric. With increasing thickness, the scaling characteristics and the multi-fractal spectrum in thicker samples gradually approach the hysteresis loop criticality in three-dimensional systems. Thin ferromagnetic films are growing in importance technologically, and our results illustrate some new features of the domain wall dynamics induced by magnetisation reversal in these systems.
\end{abstract}

\maketitle
\section{Introduction\label{sec:intro}}
Disordered ferromagnets are well-known memory materials and new
classes of memory devices are increasingly making use of
controlled motion of the domain 
walls (DW) in thin ferromagnetic films and nanowires  \cite{science-rew2008,NatNanotech2015_Tagantsev_DwFE,saveljev_NJP2005}. 
The underlying magnetisation-reversal processes in these disordered ferromagnetic materials typically exhibit domain 
nucleation and domain-wall propagation under slow driving by the
external field \cite{zapperi_PRB1998,BT_PhysA1999}.
Hence, there is an increased interest in the experimental investigations of the Barkhausen noise (BHN) accompanying 
the magnetisation reversal along the hysteresis loop in nanowires  \cite{DW_NWshape2017}, thin films
\cite{puppin2000,berger2000,bhn-PRB2005,koreans_NatPhys2007,koreans_JApplPhys2008,koreans_films2011,HL-geometryFils2011,koreans_films2013}, and systems with a finite thickness
\cite{FBohnPRE2017,BohnDurin2018SciRep}.   On the other hand, theoretical and numerical
investigations of the impact of the specific sample shape on the
magnetisation reversal processes are still in their infancy
\cite{DWpinning-notchNW,NavasVivesPRE2016,BT_shape2018,Djole2018PRE_2D-3D}. 
The domain structure in these materials is primarily related to the
intrinsic disorder that contributes to the enhanced stochasticity of the DW motion
\cite{DWstochasticity_NatComm2010,DWstochasticity_expresistanceNoise2010,DWstochasticity2011,DWstochasticity2016,MFR_BHN_experimental},
but this remains poorly understood.  

One of the key sources of the DW stochasticity are
the dynamic critical fluctuations, which have no particular scale
\cite{uwe-book2014}. These occur close to a critical disorder line
that separates two distinct dynamical regimes: on one side a weak
pinning regime with 
large propogating domains,  and on the other side a strong disorder
regime with pinned
domain walls and smaller domains. 
In this context, the changing sample shape and dimensionality can affect the
extension of the domains in one or more directions and thus alter the effects of disorder on the domain
wall propagation. Consequently, the critical disorder separating the
two dynamical regimes can vary with the sample shape and the effective
dimensionality.  More precise theoretical investigations using the
numerical studies of Ising spin model systems with the random-field magnetic disorder (RFIM)
and the concept of finite-size scaling  \cite{FSS_book} determine the
critical disorder $R_c^{3D}=2.16$ in the three-dimensional
\cite{cornell_perkovic1999,eduard_FSS2003,eduard_spanning2004}, 
and $R_C^{2D}=0.54$ in two-dimensional systems \cite{djole_PRL_critdisorder2011,djole_scaling2D}, 
augmenting earlier studies with a built-in DW
\cite{BT_PRL1996,BT_PRE2000,EQRFIM2}. 
Recently, using the extensive simulations and extending the
finite-size scaling for the systems with the base $L\times L$ and finite
thickness $l$, in Ref.\ \cite{Djole2018PRE_2D-3D} the critical disorder
line $R_c^{eff}(l,L)$ has been determined interpolating from the
two-dimensional ($l=1$) and three-dimensional ($l=L$) RFIM systems. Apart
from the value of the critical disorder, the DW motion at different
spatiotemporal scales \cite{BT_MFRbhn2016,Janicevic2018} as well as
the interplay of the critical fluctuations and the shape of the
lattice constitute the theoretically challenging problems of broad
importance. A review of some recent developments is given 
in \cite{crit-geometry}. Some of the considersations in this paper are
reminiscent of research that considers
criticality of spin systems situated on a complex network topology \cite{DD_Bethelatt1,topol2,topol1,SD-criticality}.

In this work, we tackle some of these issues through the numerical study of 
magnetisation reversal processes in  RFIM
systems of variable thickness on the critical disorder line,
moving from a two-dimensional plane to the three-dimensional lattice. 
At one end of this critical line, in the
three-dimensional limit, the hysteresis-loop behaviour was
investigated extensively by numerical methods
\cite{cornell_perkovic1999,eduard_FSS2003,eduard_spanning2004}. The
three-dimensional model is also accessible to field-theory approaches
\cite{uwe-book2014,tadicRG81,cornel-RG1996,tarjusRFIM,instantonsRFIM-Hc}. 
In the course of the reversal process along the hysteresis loop, the
occurrence of large domain walls and their motion in  the
central part of the  loop, where the external field is close to the coercive field  $H_c$,  play a crucial role in the
critical dynamics. It has been recognised \cite{instantonsRFIM-Hc}
that in the metastable states near  $H_c$ particular
configurations of the disorder can trigger a large system-wide avalanche. 
In contrast, much less is known about the structure of such states in
finite geometry 
samples or in the two-dimensional limit,  which appears to be the lower
critical dimensionality of the field-theory model. Recent numerical
investigations \cite{djole_scaling2D,BT_shape2018,NavasVivesPRE2016,Djole2018PRE_2D-3D} indicate a rich dynamical critical behaviour, prone to the impact of geometry and disorder. 
Therefore, we adopt an adiabatic driving mode, where the
field increments adjust to the current minimal local field (see
Methods) and focus on the nature of fluctuations in the central part of
the hysteresis loop.
Our analysis reveals that the samples of sufficiently small thickness
have a new critical behaviour on the hysteresis loop, which is
different from the one in 
two-dimensional limit; these differences manifest themselves at the
level of 
multifractality of the Barkhausen noise signal as well as the
avalanches of domain-wall 
slides. On the other hand,
the hysteresis-loop criticality in substantially thick samples
gradually changes with the increased thickness, increasingly
resembling the three-dimensional system.

\section{Model and Methods\label{sec:methods}}

\subsubsection{Field-driven spin reversal dynamics in RFIM}
Random-field disorder, which locally breaks the rotational
symmetry of the order parameter,   is considered to adequately describe the impact of
magnetic defects on criticality in classical \cite{belangernatterman}
and quantum \cite{quantum_ctrit2010} spin systems.
To model the effects of disorder on the magnetisation reversal along
the hysteresis loop, we use a RFIM
driven by the time-varying external field $H_{ext}$ at zero
temperature
\cite{belangernatterman,cornell_sethna1999,hysteresis_book2006} . The Hamiltonian 
of interacting  Ising spins $s_i=\pm 1$ is  
\begin{equation}
\mathcal{H}= - J\sum_{\langle i,j \rangle}s_is_j -\sum_{i}h_is_i -
H_{ext}\sum_{i}s_i \>,
\label{eq:RFIMHamiltonian}
\end{equation}
where  $i=1,2,\cdots N$ runs over all sites and $\sum_{\langle i,j \rangle}$ denotes the summation over all
pairs of nearest neighbour spins on the lattice of the specified size
$L \times L \times l$, and the strength of the ferromagnetic coupling
is fixed $J=1$.
At each lattice site, the value $h_i$ of  the random field is
chosen from the Gaussian distribution
$\rho(h)=e^{-h^2/2R^2}/\sqrt{2\pi}R$ of zero mean and the variance
$\langle h_ih_j \rangle=\delta_{i,j}R^2$. The realisation of these
random fields is considered as a quenched
disorder \cite{belangernatterman}, consequently, the fields are kept
fixed during the system's evolution. %%
The zero-temperature dynamics consists of spin-flip $s_i(t+1)=-s_i(t)$
by aligning the spin $s_i$ with its local field $h_i^{loc}$, which is given by $ h_i^{loc}= J\sum_{j}s_j+H_{ext}+h_i $. Apart from the fixed random field $h_i$ at that site, the time varying
contributions to $h_i^{loc}$ are due to the state of all neighbouring spins $s_j$ and
the actual value of the external field $H_{ext}$.  %%
The spin system is driven along the ascending branch of the hysteresis
loop starting from the uniform state $\{s_i=-1\}$ for all
lattice sites, and a large negative  $H_{ext}$. The external field
is increased for a small value to start a new avalanche (see below)
and the updated values of $h_i^{loc}$ at all sites are computed and kept
until all unstable spins are flipped in the current time step. Then the set of new
local fields  $h_i^{loc}$ is determined at sites in the shell
at the avalanche front, and the unstable spins
flipped; the process is repeated until no more unstable spins can be found. Then
the external field is increased again. Note that the number of chain
events strongly depends on the state of the system, the strength of
disorder $R$, and the actual value of the external field. 

The sequence of spin-flip events between
the two consecutive updates of the external field comprise an
\textit{avalanche}. This larger-scale event can be characterised by
the duration $T$---the number of time steps, and size $S$---the number of flipped spins during the
avalanche propagation, i.e.,
$S=\sum_{t=t_s}^{t_e}n_t $ and $T=t_e-t_s$,
where $n_t$ is the number of spins flipped during the step $t$, and
$t_s$ and $t_e$ indicate the moments when the avalanche begins and
ends, respectively. Note that in the zero-temperature dynamics the number $n_t$ gives the exact change
of the magnetisation $\delta M(t)\equiv M(t+1)-M(t) =2n_t/N$ at time
$t$, constituting the time signal known as Barkhausen noise. Here, the
magnetisation $M(t)=\sum_{i=1}^{N}s_i/N$ varies with $t$ depending on the state of all spins.
To minimally affect the avalanche propagation the
driving field is incremented  {\it adiabatically}, that is, the
external field $H_{ext}$ is held constant during each avalanche. Moreover, the field that starts a new avalanche is updated by the amount that
matches the local field of the minimally stable spin in the entire
system, which is identified using a sorted-list search method \cite{cornell_sethna1999}. The process ends
when all spins are reversed completing the hysteresis branch. We sample two sets of
systems of the size $L\times L \times l$ where the thickness $l=2^k$,
$k=0,1,2,3,\cdots $, i.e., from $l=1$ corresponding to the
two-dimensional x-y plane of the size $L\times L$ until $l=L$ complete three-dimensional sample. The linear size $L$ of the
considered systems are $L=256$ and $L=512$. The periodic boundary
conditions are applied along x-y directions while the open
boundaries are kept in the perpendicular direction of changing
thickness. For each system, we sample the number of
flipped spins $\{n_t\}$ along the entire branch of the hysteresis loop
(Barkhausen noise signal), and identify each avalanche that occurred
during the full magnetisation reversal. To complete the avalanche
statistics, we repeat the process by new samples of the random fields
with the same disorder strength $R$. The number of runs per one
set of $l$ and $L$ pair, performed at the corresponding value of effective critical disorder $R_{c}^{ eff}(l,L)$,
ranged between 500 for the large, and 60 000 for the small
samples. The sorted-lists algorithm is very efficient. For the
largest system simulated in this work, the single run time on the
Supermicro server 8047R-7RTF+ is about 5 hours.

\subsubsection{Critical disorder of systems with finite thickness}
The critical fluctuations comprise of the avalanches of all sizes
including an infinite system-size avalanche. In the finite-size $L$  systems with periodic boundary
conditions, these are represented by \textit{spanning} avalanches that occur at an effective critical 
point $R_c^{\scriptsize{\rm eff}}(L)$. 
Then
the true critical point $R_c(L\to \infty)$ is extracted by applying the finite-size
scaling collapse \cite{FSS_book}. Using these ideas and 
simulations of the RFIM in very large systems, the critical disorder has been determined as
$R_c^{3D}=2.16\pm 0.06$ in $D=3$, and  $R_c^{2D}=0.54\pm 0.06$ in
$D=2$ spatial dimensions  \cite{cornell_perkovic1999,eduard_FSS2003,eduard_spanning2004,djole_spanningaval2014}.
In the finite-size scaling spirit, a system of finite thickness
exhibits critical fluctuations for  a reduced disorder compared with the
full three-dimensional geometry. Recently, extensive simulations and the
finite-size scaling analysis of avalanches for the systems of size
$L\times L\times l$ with varied
thickness $l$ have been performed in \cite{Djole2018PRE_2D-3D}. In this case, the spanning avalanches in the x-y dimensions
are relevant in addition to the extra scaling variable $l/L$, due to
the open boundaries in the $l$-direction. The analysis led to the
critical disorder line 
\begin{equation}
R_c(l) =\frac{R_c^{3D}}{1-\Delta/l^{1/\nu_{3D}}} \ ,
\label{eq:rc_ell}
\end{equation}
where $\Delta =1-R_c^{3D}/R_c^{2D}$ and $\nu_{3D}$ is the
correlation-length exponent of the corresponding three-dimensional
system. Relevant for  this work is the effective critical disorder
of the system of the finite base length $L$ and thickness $l$ that can be
obtained from the analysis in \cite{Djole2018PRE_2D-3D}, in
particular:
\begin{equation}
\frac{R_c^{eff}(l,L) - R_c(l)}{R_c^{eff}(l,L)}= \frac{A(l)}{L^{1/\nu_{2D}}}
\label{eq:RcEffModel}
\end{equation}
where $A(l)$ was shown to scale with  the thickness as
$A(l)=\frac{(a-\Delta)l^{1/\nu_{2D}}}{l^{1/\nu_{3D}}-\Delta}$ and
$a=0.63\pm 0.18$ is the fit parameter. The respective values are
 $1/\nu_{\scriptsize  3D}=0.71$ and $1/\nu_{\scriptsize  2D}=0.19$, using
the exponent controlling the divergence of the correlation length
for $l=L$  in $D=3$,  \cite{cornell_perkovic1999},
and  $l=1$ in $D=2$ limit, \cite{djole_PRL_critdisorder2011}.

\subsubsection{Avalanche distributions and average shapes}
Regarding  the statistics of avalanches at the critical disorder, we
distinguish the loop-integrated distributions (INT),
including the avalanches that appear over the entire branch of the
hysteresis, and  the distributions (HLC) of the 
avalanches occurring only in the central part of the hysteresis loop.  In the limiting 2D and 3D cases, the distributions of the avalanche size $D(S)$ and duration $D(T)$ obey
power-law decay   $D(x,L)=Ax^{-\tau_x}\mathfrak{D}(x/L^{D_x})$ with a
finite size cut-off and corresponding fractal dimension $D_x$,
which are well studied in the literature
\cite{eduard_FSS2003,eduard_spanning2004,djole_spanningaval2014}. 
For example, for the disorder $R\geq R_c^{3D}$, the scaling function
$\mathfrak{D}_{+}$ represents a product of a polynomial and a
stretched exponential \cite{cornell_perkovic1999,hysteresis_book2006};
whereas, $\mathfrak{D}_{-}$ corresponding to
disorders $R<R_c^{3D}$ is further modified to include the spanning avalanches of different dimensions
\cite{eduard_FSS2003,eduard_spanning2004,djole_spanningaval2014}.
In the samples of  finite thickness with the lattice size $L \times L \times l$,
the appearance of the extra scaling variable $l/L$ induces substantial
changes both in the scaling function and the exponents (see
Results). In this case, we observe two distinct slopes for small and
large avalanches, respectively, which can be fitted by the following
expression 
\begin{equation}
D(S)=\Bigl \{\bigl[1-\tanh (S/B)\bigr ]\frac{A_1}{S^{\tau_1}}+\tanh
(S/B)\frac{A_2}{S^{\tau_2}}\Bigr \}\mathfrak{D}(S)\>,
\label{eq:DistribFitForm}
\end{equation}
for the avalanche size $S$ and $\mathfrak{D}(S)$ the scaling function for a particular size $L$ and thickness $l$. 
The factor in the curly brackets in (\ref{eq:DistribFitForm}) is a convex combination of
 two power-laws, $A_1/S^{\tau_1}$ and $A_2/S^{\tau_2}$, specified by the amplitudes $A_1$ and $A_2$, and
exponents $\tau_1$ and $\tau_2$, respectively. For $S\ll B$, the first power-law prevails, so $\tau_1$ gives the
slope of the log-log plot of the curve $D(S)$ in that region. Then, for $S\approx B$, the distribution curve
bends and proceeds with the second slope $\tau_2$ in the part of scaling region where $S\gg B$, up
to the large-avalanche cutoff, where the universal scaling function  $\mathfrak{D}(S)$ becomes dominant.
At disorders above the effective critical disorder,  a stretched
exponential form  $\mathfrak{D}(S)=\exp [{-(S/C)^a}]$  can be used. However, to capture the contribution of the spanning
avalanches that typically occur below the critical
disorder, a more elaborate expression for $\mathfrak{D}(S)$ is needed to distinguish between different types
of spanning avalanches, which is left out of this work.  
A similar expression (\ref{eq:DistribFitForm}) applies for  the duration $T$ of avalanches, with
the corresponding exponents $\alpha_1$ and $\alpha_2$ and a scaling
function $\mathfrak{D}(T)$. The bending value $B=S_x$ of the size and $B=T_x$ of the duration
distribution depend on the actual sample thickness (see Results).

The average size of all avalanches of given duration  $T$,  $\langle S\rangle_T$, also exhibits a scale invariance $\langle S\rangle_T\propto
T^{\gamma}$ with the exponent $\gamma=(\alpha -1)/(\tau-1)$. With two distinct
scaling regions in the distributions of size and duration, here also two
exponents $\gamma_1$ and $\gamma_2$ can be observed for some intermediate sample thicknesses.
Similarly, two values of $\gamma$ are  extracted from the data for
the average avalanche shape for small and large durations using the
analytical form \cite{Laurson2013}
\begin{equation}
\langle n_t(t\vert T)\rangle \propto T^{\gamma -1}\Biggl [\frac{t}{T}\biggl
(1-\frac{t}{T}\biggr ) \Biggr ]^{\gamma-1}\times \Biggl [1-a\biggl
(\frac{t}{T}-\frac{1}{2}\biggr ) \Biggr ].
\label{eq:AAS}
\end{equation}
Here, $\langle n_t(t\vert T)\rangle $ is the number of spins flipped  at
the moment $t$ since the start of the avalanche whose duration is $T$
and averaged over all avalanches
 of the duration $T$; the exponent $\gamma$ is defined above and $a$
 is the asymmetry parameter.

\subsubsection{Detrended multifractal analysis of Barkhausen noise signal}
As demonstrated in \cite{BT_MFRbhn2016}, the convenient approach 
of studying the  multifractal features of the  magnetisation reversal
fluctuations  exploits the underlying scale-invariance to determine the generalised Hurst exponent $H(q)$.
The respective time series $\delta  M(k)$,  $k=1,2,\cdots T_{max}$ of the length 
$T_{max}$ comprises a selected
segment of the BHN signal $\{n_t\}$ on the hysteresis loop (see Results).
Following the standard procedure described in
\cite{MFRA-uspekhi2007,DMFRA2002,dmfra-sunspot2006,BT_MFRbhn2016},
the profile of the time series
$Y(i) =\sum_{k=1}^i(\delta M(k)-\langle \delta M\rangle) $
is firstly obtained, and thereafter divided into non-overlapping segments of equal length $n$.  The
process is repeated starting from the end of the time series resulting
in total $2N_s=2{\rm int}(T_{max}/n)$ segments; here, ${\rm int}(x)$ is the integer part of a real number $x$.
  Then, the local trend $y_\mu(i)$ is found at each segment
  $\mu=1,2\cdots N_s$, which enables the determination of the standard
  deviation  $F(\mu,n)$  around  the local trend
\begin{equation}
 F(\mu,n) =\left\{ \frac{1}{n}\sum_{i=1}^n[Y((\mu-1)n+i)-y_\mu(i)]^2\right \}^{1/2} ,
\label{eq-F2}
\end{equation}
and similarly, $F(\mu,n) =\{\frac{1}{n}\sum_{i=1}^n[Y(N-(\mu-N_s)n+i)-y_\mu(i)]^2\}^{1/2}$ for $\mu
=N_s+1,\cdots 2N_s$.
Finally, the $q$-th order fluctuation function $F_q(n)$ is computed
for segment length $n$, and averaged over all segments
\begin{equation}
F_q(n)=\left\{\frac{1}{2N_s}\sum_{\mu=1}^{2N_s} \left[F^2(\mu,n)\right]^{q/2}\right\}^{1/q} \sim n^{H(q)}  \ .
\label{eq-FqHq}
\end{equation}
The idea behind this formula is that various segments of the
signal need to be enhanced in different ways (values of $q$) to achieve
a self-similarity of the whole signal. In particular,  \textit{small
  fluctuation segments} are enhanced by the
negative values of $q$, while the segments with \textit{large
  fluctuations} dominate the fluctuation function  for the
positive values of $q$.
By varying the  segment lengths in the range $n\in[2,{\rm int}(T_{max}/4)]$,
we compute the fluctuation function $F_q(n)$ for different $q\in
[-10,10]$.   Plotting $F_q(n)$  against $n$
allows us to find the regions of scale
invariance  and the corresponding scaling exponent $H(q)$, as the
slope of straight lines in the double-logarithmic plot.
 Furthermore, the exponent $\tau (q)$ of the box
probability measure,  standardly defined in the partition function method, is related to
$H(q)$ via the scaling relation $\tau (q)=qH(q)-1$.  Hence, the
singularity spectrum  $\Psi(\alpha)$ is obtained from $H(q)$ via the
Legendre transform  of $\tau (q)$. In
particular,  $\Psi(\alpha)=q\alpha -\tau (q)$, where $\alpha =d\tau/dq
= H(q)+qdH/dq$. For a monofractal, we have $H(q)=H=const$ and $\alpha =
H$; consequently,  $\Psi (\alpha)$ reduces to a single point.

\section{Results\label{sec:results}}
\subsubsection{Hysteresis loop and signal shape in thin samples at the
  critical disorder}
\begin{figure}[!htb]
\includegraphics[width=0.9\linewidth, trim=0.0cm 1cm 0.0cm 0.00cm,clip=true]{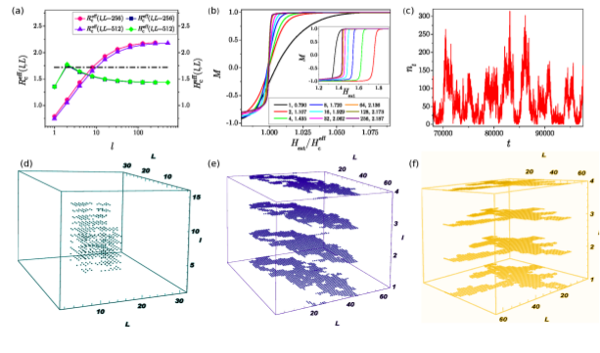}
\caption{\small  
(a) Effective critical disorder $R_{c}^{eff}(l,L)$ and the coercive field 
$H_{c}^{eff}(l,L)$ versus sample thickness $l$, for
the systems with base size $L=256$ and $L=512$. 
The horizontal line at a fixed disorder is to  indicate a typical
variation of system parameters accessible to experiments (see Supplementary Information). 
(b) Magnetization $M$ against the rescaled magnetic 
field $H/H_{c}^{eff}$ for various sample thicknesses $l$ and base size $L=256$; 
for each $l$ the magnetization curve is obtained at the corresponding  effective critical 
disorder $R_{c}^{eff}(l,L)$ shown in the legend.
 Inset: the same magnetization curves versus the magnetic field $H$. 
(c) An example of the BHN signal $n_t$ against time $t$; the 
fragment  is extracted from the response of a system at the critical disorder 
$R_{c}^{eff}$ for $L=256$ and small thickness $ l=4$.
(d-f) Sample avalanches:  non-spanning ($R=2.5,\>l=16,\>L=32$), 
1D-spanning ($R=1.9,\>l=4,\>L=64$), and  2D-spanning
($R=1.8,\>l=4,\>L=64$), respectively.
}
\label{fig-forFig1_Rceff_}
\end{figure}

  The critical disorder line $R_{c}^{eff}(l,L)$ for the sample with
  the base size $L=256$
and varied thickness $l=2^k$, $k=0,1,2,\cdots 8$ is plotted in Fig.\ \ref{fig-forFig1_Rceff_}a together with the  effective coercive field
$H_{c}^{eff}(l,L)$; the corresponding lines for the case of $L=512$
are also shown.
The effective critical disorder increases at small thickness and then
saturates approaching the values for the 3D samples.  
As expected in disordered materials \cite{belangernatterman,HL-geometryFils2011}, the increased disorder
induces narrowing of the hysteresis loop, which is compatible with
the smaller values of the effective coercive fields  
$H_{c}^{eff}(l)$ for  $l\geq 2$,
as shown in Fig.\ \ref{fig-forFig1_Rceff_}b. 

The small thickness also affects the shape of the signal and the propagation of
avalanches, as demonstrated in Fig.\ \ref{fig-forFig1_Rceff_}c-f. As
mentioned above, the avalanches of different sizes including the
sample-spanning avalanches are expected at critical
disorder. In the case of small thickness, the avalanche often hits
the system's open boundary in the z-direction and stops, while the propagation
in the x-y directions within the sample is conditioned by the pinning
of avalanches by the random-field disorder; some examples of
avalanches are shown in Fig.\ \ref{fig-forFig1_Rceff_}e-f. 
Hence, the sample thickness determines the actual shape of the
critical avalanches. 
These effects are also manifested in
the shape of the accompanying BHN signal. For example, for a large
and thin sample, see Fig.\ \ref{fig-forFig1_Rceff_}c,
small variations of the signal occurring due to pinning at the
boundary appear intermittently between the
large fluctuations even in the central part of the hysteresis
loop. A detailed analysis below reveals how these  fluctuations are  manifested in the multi-fractal
properties of the BHN signal in the thin samples.
\begin{figure}[htb]
\begin{tabular}{cc}
\resizebox{20pc}{!}{\includegraphics{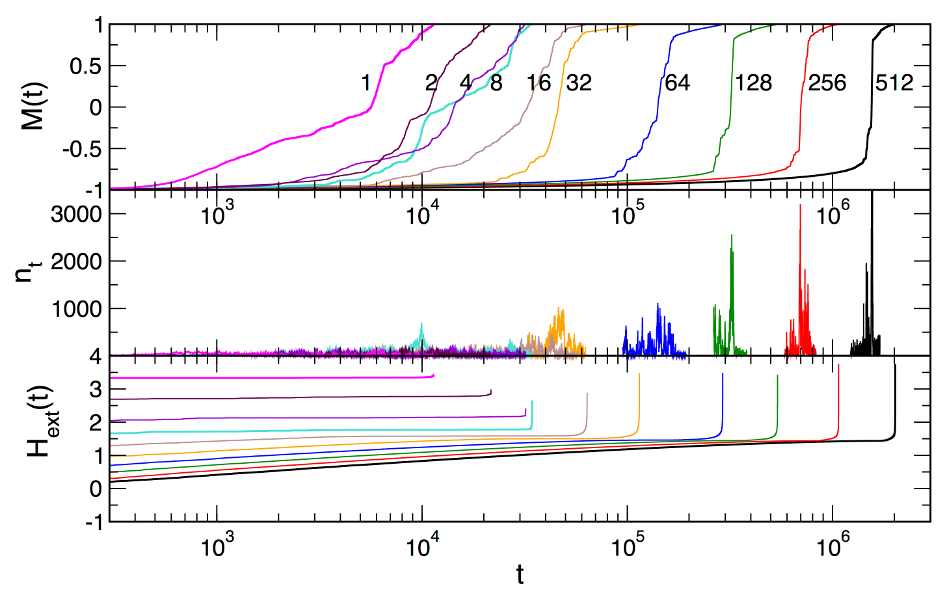}}\\
\end{tabular}
\caption{\small Magnetisation $M(t)$ plotted against  time $t$ (top panel), the
  corresponding BHN signal $n_t$ (middle panel), and the time-varying
  external field $H_{ext}(t)$
  (lower panel) for the increasing sample thickness $l$, indicated
  in the top panel, from 2D sample  $l=1$ to 3D sample
$l=L=512$. The
  part of the signal corresponding to the center segment of the
  hysteresis loop are shown in the middle  panel. The logarithmic
  scale along the time axis is applied. The beginning of the
loop is omitted to improve clarity. }
\label{fig-Mt-nt_d}
\end{figure}

 Fig.\ \ref{fig-Mt-nt_d} shows how the sample thickness affects the magnetization 
increase with time in the ascending branch of
the hysteresis loop. Precisely, the pronounced effects  occur 
in the case of small thickness 
$l \lesssim l_{tr}$,  where $l_{tr}\approx L/8$ is a transitional
thickness, which depends on the 
 base size $L$. In contrast,  for the thicker samples with $l > l_{tr}$,
the effects of the finite thickness are more predictable, as the
analysis below will show. The majority of the critical fluctuations
permitting the 
spanning avalanches occur in the central part of the
hysteresis loop (HLC); therefore,  we mainly focus on these segments of the
loop. The corresponding segments of the BHN signal at each sample thickness
are indicated  in the middle panel of Fig.\ \ref{fig-Mt-nt_d}, while the
related values of the external fields that cause these
fluctuations are given in the lower panel.
Note that, due to the adiabatic driving where the field is kept fixed during the
avalanche propagation, the effective driving rate in the HLC segments is
minimal, thus allowing a spontaneous evolution of the system.
\begin{figure}[!htb]
\includegraphics[width=0.9\linewidth, trim=0.0cm 0cm 0.0cm 0.00cm,clip=true]{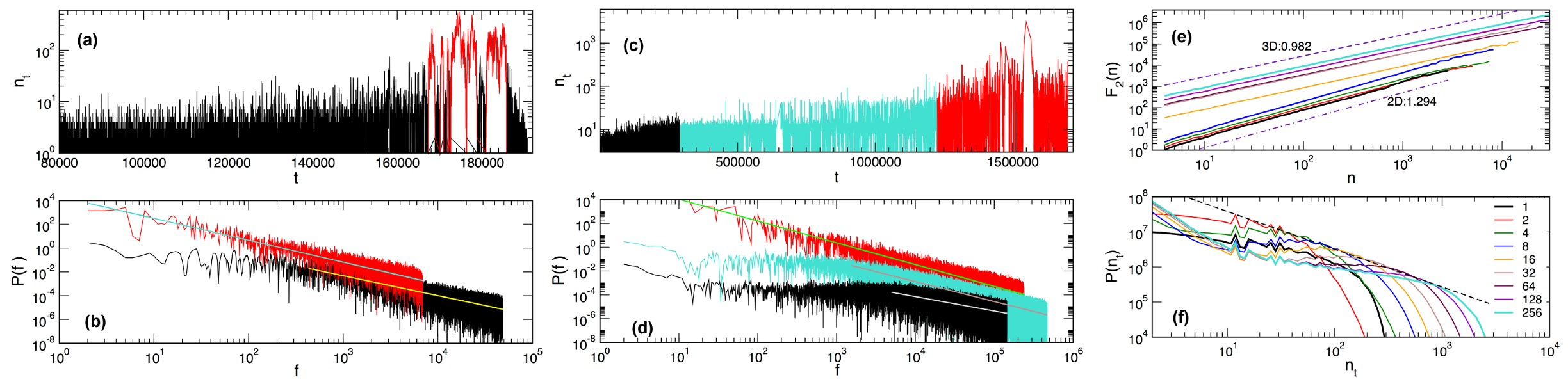}
\caption{\small  Signal selection according to the avalanche sizes for $L=256$, $l=32$ (a,b), 
and according to three hysteresis-loop segments for $l=L=512$ (c,d). The lower panel in each 
case shows the corresponding power spectrum of the selected signals against frequency $f$ with the slopes 
$\phi= 1.84\pm 0.02$ and $1.67\pm 0.02$, panel (b), and $\phi= 1.835\pm 0.014$, 
$1.673\pm 0.008$, and $1.208\pm 0.009$, panel (d).
%Hurst exponent $H(2)$ 
Second-order fluctuation function $F_2(n)$ vs segment length $n$  
for $L=512$ and varied thickness $l=512$ top line to $l=1$ bottom line (e); 
the two dashed lines have the slopes equal to the Hurst exponent $H(2)$ in 
2D and 3D case. Distributions of the height $n_{t}$ of the BHN signal for
$L=256$ and varied  thickness $l$  indicated in the legend, and a
tangent---dashed line, (f).}
\label{fig-nt_PS_segments}
\end{figure}

As mentioned above, the size and thickness of the sample affects the
critical disorder $R_{c}^{eff}(l,L)$ and, consequently, the shape of the BHN signal.
Some features of the BHN signal obtained at the critical disorder in samples of different
thickness  are illustrated in Fig.\ \ref{fig-nt_PS_segments}a-f. The
signals exhibit long-range temporal correlations with the power
spectrum $P(f) \sim f^{-\phi}$ over an extended range of frequencies $f$.  
The previous studies of the multifractal features of the BHN in 
2D and bulk samples in a strong disorder regime
\cite{BT_shape2018,BT_MFRbhn2016}, suggest that the signal shape
differs in 
different segments of the hysteresis loop. Here, we
demonstrate how the size and temporal correlations of the signal
change  along the 
hysteresis loop in the 3D sample
at the critical disorder, see Fig.\ \ref{fig-nt_PS_segments}c-d. 
Moreover, in the present context, it is interesting to point out another
segmentation of the signal, which comprises  the separation of small and
large avalanches  occurring in a thin sample. Two panels in Fig.\ \ref{fig-nt_PS_segments}a-b show the  
respective separation and the corresponding power spectra for  
a sample of  the transitional thickness $l_{tr}(L)$ for the given base size $L$ 
(see below for its precise definition).
Furthermore, the persistent fluctuations are observed that are compatible with the Hurst exponent
$H(2) \lesssim 1$ in samples of a larger thickness, whereas $H(2) > 1$
for the thin samples having 
$l <l_{tr}(L)$, as shown in Fig.\ \ref{fig-nt_PS_segments}e.   For the analysis in this paper, it is
also important to note that the distribution of the signal heights (data points)  
show a  broad peak  
that moves to the right with the increased
sample thickness, as shown in Fig.\ \ref{fig-nt_PS_segments}f. The
tangent line has a power-law  
slope, while the small signal heights
have entirely different distribution, which also varies with the thickness.

\subsubsection{Critical avalanches in samples of different thickness}
\begin{figure}[!htb]
\includegraphics[width=0.8\linewidth, trim=0.0cm 0cm 0.0cm 0.00cm,clip=true]{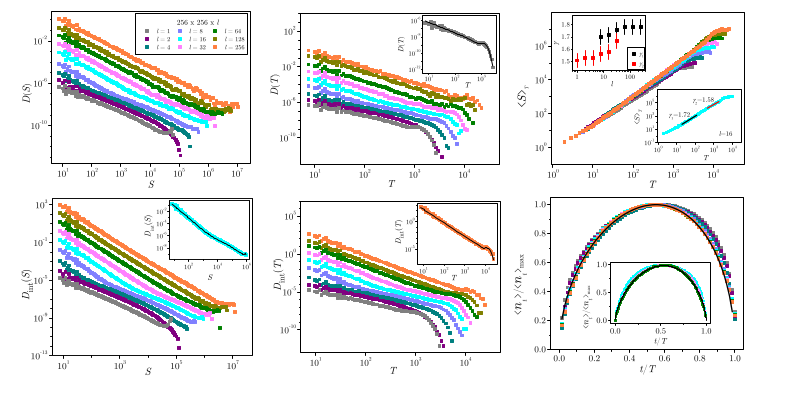}
\caption{\small Avalanche distributions for different thicknesses
  $l$ given in the legend (applies to all four panels).
Left column: size distributions in the HLC (top), and loop-integrated
size distributions (bottom panel, with the best fit of type
(\ref{eq:DistribFitForm}) for $l=16$ in the inset).
Middle column: corresponding duration distributions (top  panel, with
the best fit of type (\ref{eq:DistribFitForm}) for $l=1$ in the inset), and 
integrated duration distributions  (bottom  panel, with the best fit of type (\ref{eq:DistribFitForm}) for $l=256$ in the inset).
Right column: average size of avalanches $\langle S\rangle_T$ having a
particular duration $T$; Insets: determination of the
exponents $\gamma_1$ and $\gamma_2$ (lower-right), and their variation
with $l$ (top-left inset).
Bottom right panel shows the normalised average avalanche shapes
$\langle n_t\rangle$ vs $t/T$ for various $l$ and the fixed duration
$T=64$ (main panel) and  $T=2048$ (inset). Fits according to (\ref{eq:AAS})
with $a$=-0.214, $\gamma$=1.51, main panel, and $a$=-0.176,
$\gamma$=1.628, inset.}
\label{fig:DSDTAAS}
\end{figure}

In Fig. \ref{fig:DSDTAAS} we show the distributions of size $D(S)$ and
duration $D(T)$ of the avalanches obtained for various  sample
thicknesses. These distributions contain avalanches collected in
the central part of the hysteresis loop in a small window of the external
field, and are most relevant for the critical dynamics. In addition,
we also show the results for the avalanches collected along the entire
hysteresis branch, denoted by $D_{int}(S)$ and $D_{int}(T)$, for size
and duration, respectively, that are typically
determined in the analysis of the experimental BHN signals. 

In contrast to the avalanches in strictly two-dimensional
\cite{djole_PRL_critdisorder2011,djole_scaling2D} and 
three-dimensional RFIM 
\cite{eduard_FSS2003,eduard_spanning2004,hysteresis_book2006}, the
avalanche distributions in the samples of  finite thickness
exhibit two distinct scaling regions, for small and large avalanches,
respectively, as shown in Fig.\
\ref{fig:DSDTAAS}. More specifically, the first larger slope
(identified by the exponent $\tau_1$ and similarly $\alpha _1$, see
Methods) describes the scale-invariant behaviour of small
avalanches. Whereas the second region with a smaller exponent $\tau_2$
(and $\alpha _2$) relates to the avalanches larger than the bending
size $S_x$ (or duration $T_x$). The bending point $S_x$ (and
corresponding $T_x$) depends on the
sample thickness and the base size, and gradually moves towards larger
size  with an increased sample thickness. 
Thus, we find that the larger slope appears and  can be measured for
the lattices  of quite small thickness; it  gradually wins, and when
$l\to L$  approaches  the exponent $\tau_1\to \tau^{3D}$ and $\alpha_1\to
\alpha ^{3D}$ (note open boundary conditions). 
The two-slope distributions are typically found for sufficiently thin
samples, i.e.,  $8\le l\le 32$ for $L=256$, and $16\le l\le 64$ for $L=512$ (see also Fig.\
\ref{fig-nt_PS_segments}e). It should be stressed that these features
apply to both the loop-integrated avalanches as well as the
avalanches in the central part of the hysteresis, as also demonstrated
in Fig.\ \ref{fig:DSDTAAS}. Therefore, although the corresponding
exponents are somewhat smaller in the central hysteresis segment, the occurrence of two scaling
regions in the distributions of the critical Barkhausen avalanches is
a unique property of the thin samples with $l\lesssim l_{tr}$. According to these
results (see also the discussion on multifractality below), the transient thickness can be estimated as $l_{tr}\approx L/8$
above which the system effectively behaves as a thick sample.

Moreover, our findings indicate that the bending size scales as 
 $S_x\propto l^{D_f}$, where $D_f=2.78$ is the fractal dimension of
 nonspanning avalanches in three  dimensions
\cite{eduard_FSS2003,eduard_spanning2004,hysteresis_book2006}. Similarly,
for the  
duration distributions, the bending duration $T_x\propto l^{z_d}$, where $z_d=1.7$ is the dynamical critical exponent of the 3D model, describing the scaling of the avalanche's duration with the linear size.
Using the expression (\ref{eq:DistribFitForm}) proposed in Methods for
the  avalanche distributions, we estimated the two sets of
exponents for the power-law region of the distribution of avalanche
size, and the corresponding exponents of the avalanche duration for samples of different thickness. 
The estimated values of the exponents $\tau_1$ and $\alpha_1$ and $\tau_2$
and $\alpha_2$ are summarised in Table-I in the Supplementary Information
both for the distributions in the central hysteresis part and the
loop-integrated distributions. The appearance of two
scaling regions in the avalanche distributions manifests in the plots
of the average size $\langle S\rangle _T$ of avalanches having duration $T$, 
shown in the right column of Fig. \ref{fig:DSDTAAS} (top panel), and
the average avalanche shapes, (bottom panel). The corresponding
scaling exponent $\gamma$, defined via  $\langle S\rangle _T
\sim T^\gamma$ (see Methods), 
also appears to have distinct values $\gamma_1$ for small,
and $\gamma_2$ for large avalanches at the intermediate thickness $l\le
l_{tr}$, see the top inset. 
Within the error bars, the estimated values fall in the range
$\gamma_1=1.73\pm 0.05$ for large $l$,  which agrees with the value found in the case
of the equal sized 3D cubic lattices
($\gamma_{3D}=1.73$), while the values for $\gamma_2=1.56\pm 0.06$ are lower,
and close to the case in 2D square lattices
($\gamma_{2D}=1.55$).
The average avalanche shapes collected from all
sample thicknesses appear to be asymmetric, see the lower-right
panel of Fig. \ref{fig:DSDTAAS}.  The longer avalanches appear to be
more symmetrical and the value for $\gamma$ estimated  using the expression
(\ref{eq:AAS})  is bigger compared with the shape parameters of the short avalanches.

\subsubsection{Multiscale multifractality of the critical BHN signal}
\begin{figure}[htb]
\includegraphics[width=0.8\linewidth, trim=0.0cm 0cm 0.0cm 0.00cm,clip=true]{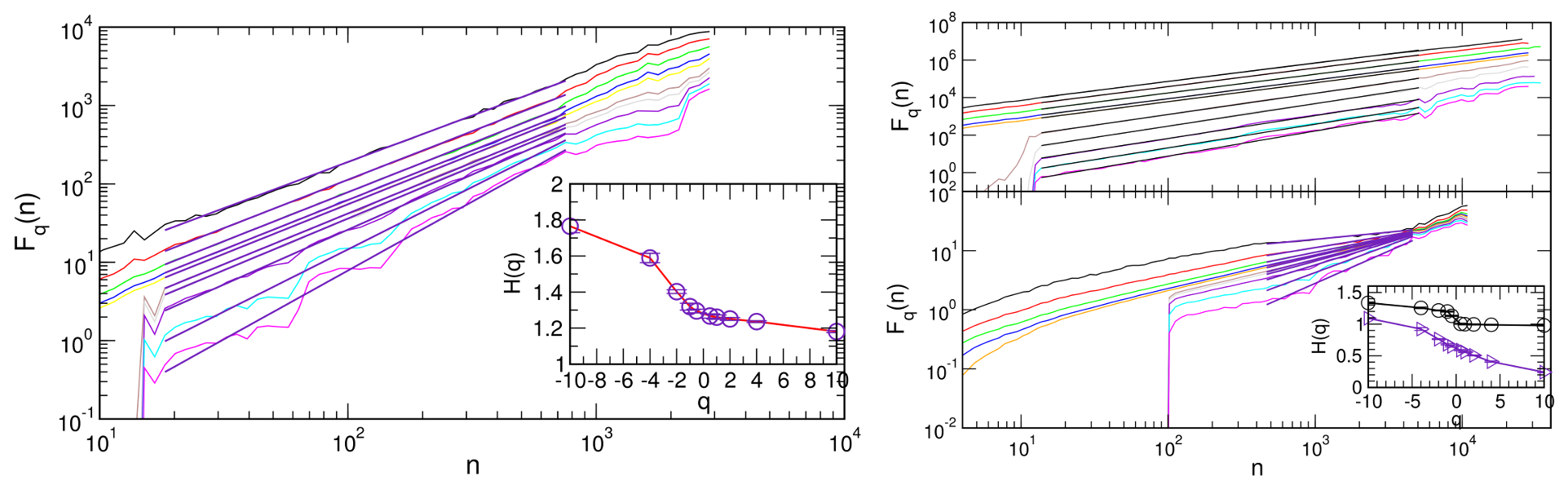}
\caption{\small Fluctuation function $F_q(n)$ for different values of $q \in [-10,10]$ for the whole signal for the 2D sample of size
  $L=512$ (left panel), and 3D sample in the HLC (right, upper panel) and 
the initial segment of the loop 
(right, lower panel). Insets: corresponding generalised Hurst 
exponents $H(q)$ against the  amplification parameter $q$, see text.}
\label{fig-Fq_2D-3D_L512}
\end{figure}

The properties of BHN signal at different sample thicknesses in Fig.\ \ref{fig-nt_PS_segments}e 
suggest that the magnetisation fluctuations are persistent with
the (standard deviation) Hurst exponent varying between 
$H(2)\lesssim 1$ in 3D samples to 
$H(2) \approx 1.29$ in the 2D case. To understand the
impact of the sample thickness on  the multifractal features of BHN  
signals, we first analyse the two limiting cases. The fluctuation
function $F_q(n)$, defined by (\ref{eq-FqHq}) in Methods,  is computed for the samples of
size $L=512$ with $l=1$ (2D sample) and $l=L$  (3D sample),  
and shown in Fig.\ \ref{fig-Fq_2D-3D_L512}. 

In the 2D limit, the scale invariance of the fluctuation function
$F_q(n)$, c.f. left panel in Fig.\ \ref{fig-Fq_2D-3D_L512}, shows that the whole
signal exhibits multifractal properties 
for a broad  range of time scales $n$ with the generalised Hurst 
exponents $H(q)\in [1.2,1.8 ]$, shown in the inset.  For
the 3D case, however, the signal in different segments of the
hysteresis loop exhibits different features, see also Fig.\
\ref{fig-nt_PS_segments}c-d for the signal segments and their power spectra. Specifically, in agreement with previous studies 
\cite{BT_MFRbhn2016}, the signal in the central segment of the
loop has $H(q)>1$, while the fluctuations at the very beginning of the loop
are governed by the actual random field distribution, resulting in a
fractional Gaussian type of noise (fGN);  consequently, its multifractal spectrum
remains in the range $H(q)<1$. For both
cases  the exponents $H(q)$ are shown in the inset of the right panel of Fig.\ \ref{fig-Fq_2D-3D_L512}.
As stated above, here we focus on the impact of thickness on
the critical fluctuations, which are prominent in the central part of
the hysteresis loop. A systematic analysis of the entire hysteresis loop
for a particular sample shape is left for another study.

\begin{figure}[htb]
\includegraphics[width=0.8\linewidth, trim=0.0cm 0cm 0.0cm 0.00cm,clip=true]{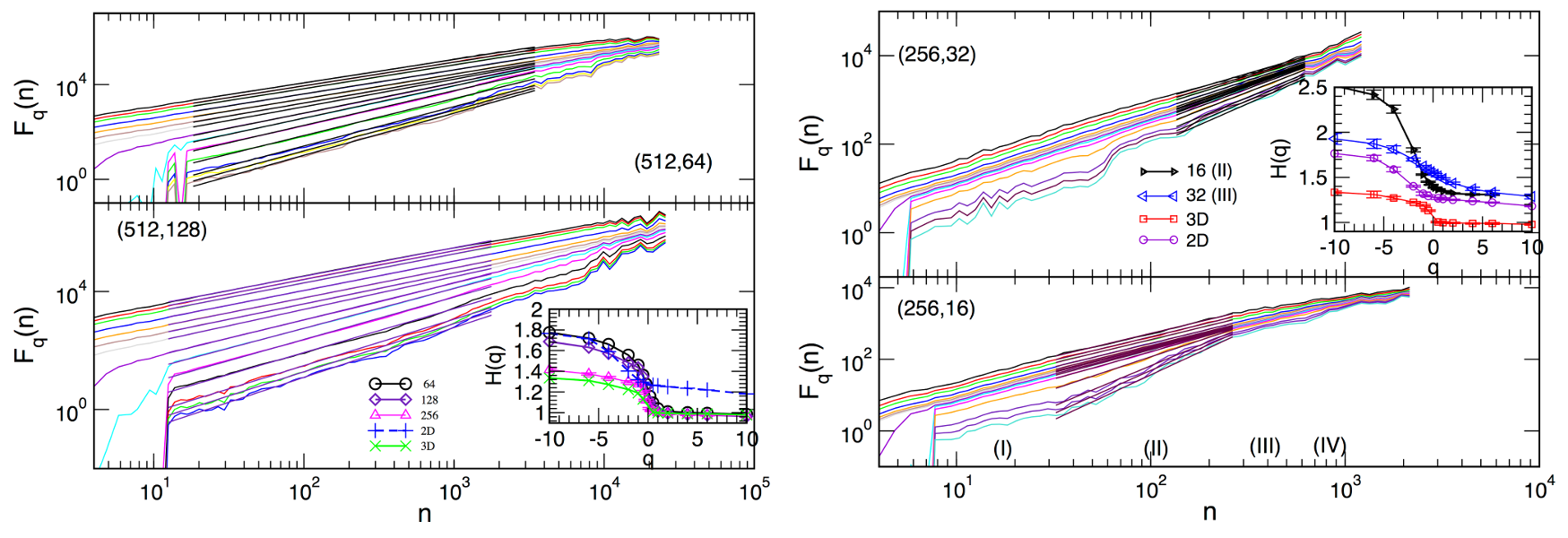}
\caption{\small  Fluctuation function $F_q(n)$ for $q\in[-10,10]$ computed in the central part of the 
hysteresis loop for samples of different base size $L$ and thickness
$l$; each pair $(L,l)$ is indicated in the corresponding panel.  
Straight lines indicate the fitted scaling regions, and the corresponding generalised Hurst exponents 
$H(q)$ are plotted against $q$ in the insets. See text for more details.}
\label{fig-Fq_ell_2x}
\end{figure}

In the finite samples, 
the spectrum $H(q)$ changes,  depending on the ratio 
$l/L$  of the thickness $l$ relative to the base size $L$ of the
system, see Fig.\ \ref{fig-Fq_ell_2x}.  Our numerical analysis  suggests 
that the most dramatic  changes occur in small fluctuations region ($q<0$) and 
when the samples are sufficiently thin such that 
$l/L\lesssim 1/16$. 
More specifically, for relatively thick samples with 
$l/L\geq 1/8$, left panels in Fig.\ \ref{fig-Fq_ell_2x} 
show that the multifractal features are apparent in a broad range of time scales $n$. 
For $q>0$, the exponents $H(q)$ remain in the area of the standard Hurst exponent 
$H(2)$ of a 3D sample, whereas 
significant deviations occur in $q<0$ region, governing
small fluctuations. This part of spectrum gradually approaches the
one observed in the bulk 3D samples when  
$ l \to L$, as shown in the inset.

On the other hand, the thin samples with $l/L\lesssim 1/16$   
exhibit a time-scale dependent behaviour of the fluctuation function $F_q(n)$, 
c.f. right panels in Fig.\ \ref{fig-Fq_ell_2x}. 
Here, we find that several scaling regions occur, indicated by
(I)-(IV) in the lower right panel, where different spectrum $H(q)$ can be determined. 
While the multifractality in some of these
regions is apparent (see, for example, the region (II) in 
Fig.\ \ref{fig-Fq_ell_2x}), some of the other
areas appear to have a narrow spectrum which is virtually
monofractal, see, for example region (I).   In all cases, the values of the generalised Hurst
exponents, shown in the inset,  are in the range above the corresponding  values 
in the 2D limit. Again, the most significant deviations occur in the negative part of the 
spectrum $q<0$. See further discussion in the next section.

\begin{figure}[htb]
\includegraphics[width=0.8\linewidth, trim=0.0cm 0cm 0.0cm 0.00cm,clip=true]{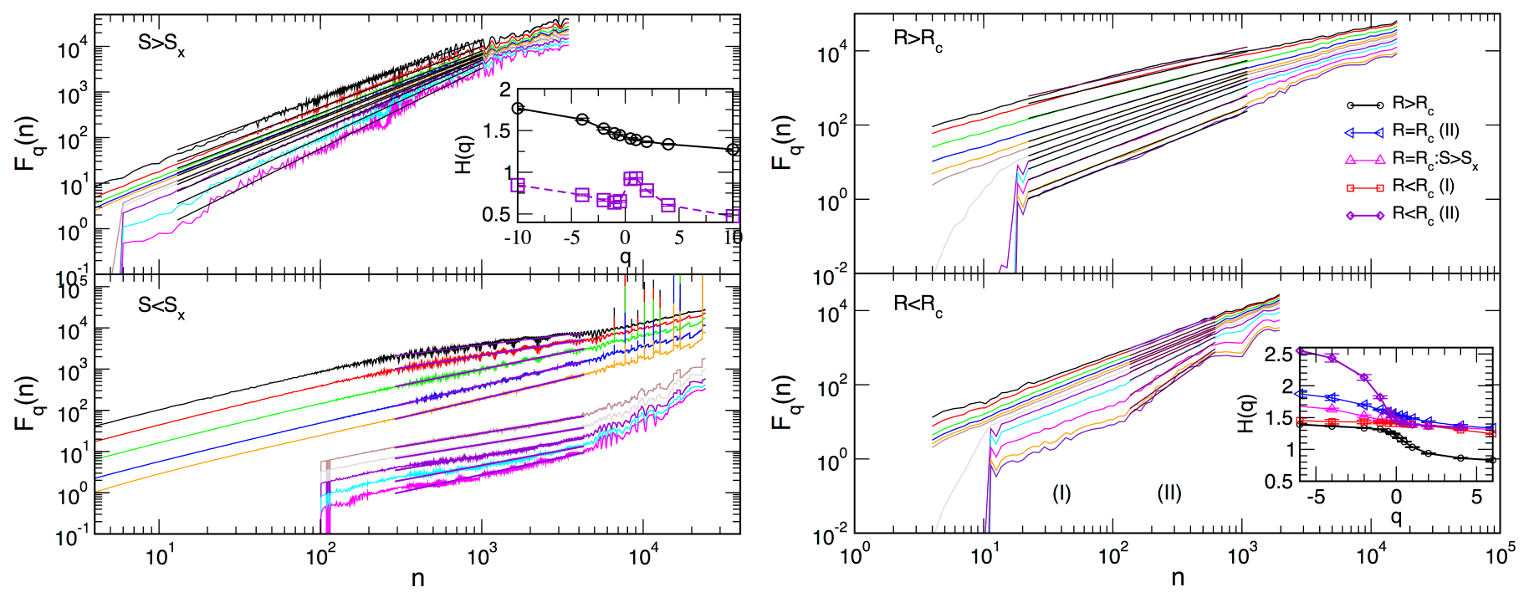}
\caption{\small For the sample of base size $L=256$ and thickness $l=32$,  the fluctuation 
function $F_q(n)$ of the signal segments selected according to the avalanches above 
(below) the bending size $S_x$, left panels, and signals in the HLC for the disorder above (below) the effective critical disorder, right panels, denoted as $R_{c}$. 
Corresponding generalised Hurst exponents $H(q)$  are shown against $q$ in the respective insets.}
\label{fig-Fq_avsel32_R}
\end{figure}

\section{Discussion and conclusions\label{sec:discussion}}

Our study of the magnetisation reversal processes \textit{along the
effective critical disorder line} $R_{c}^{\scriptsize{\rm eff}}(l,L)$ revealed  that the pinning of DW at the open boundaries in the direction of
small thickness can hinder the propagation of avalanches and the shape
of the BHN signal at all scales. For example, in the sufficiently thin
samples, many small avalanches can occur whose scaling exponents are close to the 3D
RFIM class, at the same time, large avalanches with the size
above the bending size $S_x(l)$, manage to propagate
in the two transverse directions, resembling quasi-2D avalanches.
The relative contribution of these  two types of avalanches to the whole process
gradually changes as the sample thickness  $l$ increases.
Consequently, the fluctuations of the magnetisation in the central part
of the hysteresis loop changes with the increased thickness.
More specifically, for the thick samples,  $l \gtrsim l_{tr}\sim  L/8$, the fractality of the significant fluctuations ($q>0$) in the 
HLC virtually coincides with the spectrum of 3D samples, whereas the
part of the spectrum with the dominant small fluctuations ($q<0$)
varies, interpolating from the 2D to the 3D case with increasing thickness. 
On the other hand, in the non-central parts of the hysteresis loop
(excluding the very beginning, where all signals are fGN type) and
along the whole hysteresis branch of thin samples  ($l <l_{tr}$), the
large and small avalanches intermittently occur, leading to a more
complex behaviour of the fluctuation function.  Consequently, the
generalised Hurst exponent dependence on the time scale (interval
length) can be observed. Interestingly, 
the intervals where multifractal features are
apparent roughly coincide with the parts of the signal that are
dominated by the 
large quasi-2D avalanches (see Fig.\ \ref{fig-Fq_avsel32_R} and the
discussion below); in both cases, the generalised Hurst exponents $H(q) >1$, and is close to the 2D sample spectrum.

\begin{figure}[htb]
\includegraphics[width=0.6\linewidth, trim=0.0cm 0cm 0.0cm 0.00cm,clip=true]{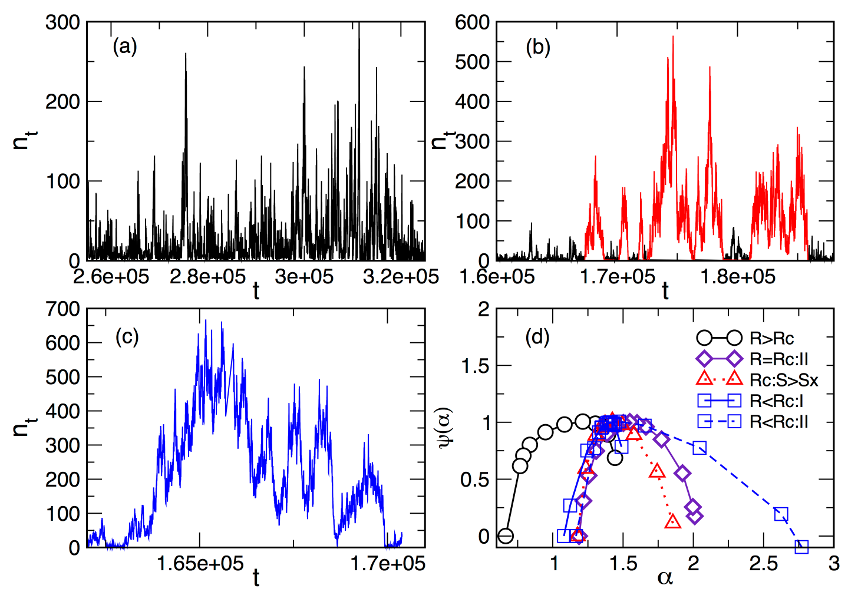}
\caption{\small (a-c) Noise signal in the central part of HL for $R>R_{c}$,
  $R=R_{c}$, and $R<R_{c}$ at the transitional sample thickness $l_{tr}=32$ for
  $L=256$ and the corresponding effective critical disorder, for simplicity denoted  by $R_c$. 
(d) The singularity spectra corresponding to these signals
  and indicated interval range, together with the spectrum referring to large
  avalanches selection, $S>S_x$. }
\label{fig-Nt_Psi_d32_R}
\end{figure}

To further support these findings, we selected the segments of the
signal that correspond to large (small) avalanches, where the bending
point $S_x(l)$ is taken from the corresponding distribution of avalanche sizes,  c.f. Fig.\ \ref{fig:DSDTAAS}.  An example of the signal selection
for $L=256$ and   $l=32$ in shown  in Fig.\ \ref{fig-Nt_Psi_d32_R}b. The fluctuation function corresponding to the separate analysis of these
parts of the signal is given in left panels of Fig.\
\ref{fig-Fq_avsel32_R}. The large avalanches, which mostly
occur in the center of the hysteresis loop, contribute to the
leading multifractal spectrum with $H(q)>1$, see inset in Fig.\
\ref{fig-Fq_avsel32_R} for $S>S_x$. Small avalanches, however, exhibit
more complex behaviour resulting in several regions with different
scaling of the fluctuation function. For instance,  $F_q(n)$ in the
intermediate-scale region, marked in Fig.\ \ref{fig-Fq_avsel32_R} for
$S<S_x$, shows different slopes than the two adjacent
regions. Moreover, the smaller slopes of the curves for $q<0$ compared
to $q>0$ results in a non-smooth spectrum $H(q)$, also shown in the
inset above.  Thus, the number of small avalanches that occur due to
pinning of the domain walls at the open boundary in samples of small
thickness can lead to the observed  multi-scale multifractality of these signals.

Next, we investigate whether these features of the BHN signal are exclusively related to
the critical avalanches. We perform simulations of the magnetisation reversal
in several disorders $R\neq R_{c}^{eff}(l,L)$ sightly above the effective critical disorder,  $R>R_{c}^{eff}(l,L)$, and slightly below it. The
corresponding fluctuation functions $F_q(n)$ are given
in the right panels of Fig.\ \ref{fig-Fq_avsel32_R} for the sample of
transitional thickness $l_{tr}/L=1/8$ and $L=256$.
 The related signal shapes and the singularity spectra $\Psi (\alpha)$
are given in Fig.\ \ref{fig-Nt_Psi_d32_R}.  While the relative size of
the time scale changes compared to the critical fluctuations, the
scale-dependent multifractality of the signal clearly
persists for stronger disorder $R\gtrsim R_{c}^{eff}(l_{tr},L)$.
Here, although all avalanches are smaller than the ones at the critical disorder, the co-occurrence of small and large avalanches can be
distinguished both in the signal, see Fig.\ \ref{fig-Nt_Psi_d32_R}a, and in the
avalanche distributions (see Supplementary Information, Fig.\ S2). 
Below $R_{c}^{eff}(l,L)$, however, the
  extended range of the time intervals with virtually monofractal
  behaviour appears, region (I), shifting the range of the apparent MFR towards larger
  time scale, region (II), c.f. Fig.\ \ref{fig-Fq_avsel32_R} lower right panel. At such disorders, 
  a huge avalanche of a prolonged duration appears, as shown in Fig.\ \ref{fig-Nt_Psi_d32_R}c, whose shape differs from
 the typical sharply-cut avalanches seen in the case of periodic
  boundaries that allow depinning of a DW. It is
  interesting to note that the left part of the singularity spectrum
  $\Psi (\alpha)$, which is associated with
  the large magnetisation fluctuations in this signal, coincides with
  the corresponding spectrum of the critical fluctuations and its
  part containing the selected large avalanches.   The right parts of
  these spectra, representing small fluctuations (negative $q$) are
  different in each of these cases, see Fig.\
  \ref{fig-Nt_Psi_d32_R}d. Compared to these, the spectrum
  $\Psi(\alpha)$ for the case $R>R_{c}^{eff}(l,L)$ is 
shifted towards the smaller  values $\alpha <1$, influenced by the fGN
  signal in the strong-disorder regime.

As mentioned earlier, there is considerable interest in  the
experimental investigations of Barkhausen
noise in thin films and samples of different thickness, e.g.
\cite{puppin2000,bhn-PRB2005,koreans_NatPhys2007,koreans_JApplPhys2008,koreans_films2011,HL-geometryFils2011,koreans_films2013,FBohnPRE2017,BohnDurin2018SciRep}. The
behaviour of Barkhausen avalanches observed in these systems
depends on the sample composition,
driving mode, and the segment of the hysteresis loop where the analysed
signal originates, as well as the sample thickness. The alloys ${\mathrm Ni_xFe_{1-x}}$  are often studied with a fixed
$x\approx 0.8$ \cite{bhn-PRB2005,FBohnPRE2017,BohnDurin2018SciRep} or variable $x\in [0,0.5]$ composition \cite{koreans_JApplPhys2008,koreans_films2011}
as a good system where the properties of Barkhausen
avalanches can be changed by varying the thickness and
composition. 
It should be noted that in contrast to the adiabatic driving used in
the numerical 
investigations, the experimental studies, for example in 
\cite{BohnDurin2018SciRep,FBohnPRE2017}, are performed with a
\textit{finite sweep rate} of the applied magnetic field. Moreover,
samples of various thicknesses are prepared by the same
method and, presumably, have some  \textit{constant disorder}, which
is difficult to quantify, but will presumably depend on the composition and
type of the alloy. In contrast, methods for modifying the disorder \cite{berger2000} are developed in
\cite{koreans_JApplPhys2008,koreans_films2011} for films of a
constant small thickness, and the domain walls are
directly monitored in response to a fixed field. 

If we suppose that the RFIM captures the scaling features of the Barkhausen
avalanches in these disordered ferromagnets,  it is tempting to consider the
available experimental results in view of our numerical investigation. 
In the theoretical phase diagram  showing the critical line of the effective RFIM disorder at a finite
thickness $l$ relative to the bulk sample, $R_{c}^{eff}(l,L)/R_{c}^{3D}$
vs $l/l_{ref}$, see Fig.\ S3 in Supplementary Information (SI), the above
mentioned experimental situations comprise of a horizontal line at a
fixed disorder, or a vertical line at a fixed thickness. Each of these lines
intersects with the critical disorder line at a particular point, as
illustrated in Fig.S3. Theoretically, the
\textit{change of the scaling behaviour occurs at the point where the
  critical line is crossed}.
Thus, for the thin samples left (above) the critical line, the actual disorder appears to be stronger
than the critical, while in the thicker samples on the right (below)  the critical line, the disorder
is weak, permitting large system-size avalanches due to DW depinning
that may occur in the inner part of the hysteresis loop. 
A more detailed \textit{comparison of the avalanche exponents} measured in the
crystalline samples $\mathrm{Ni_{0.79}Fe_{0.21}}$ of different
thickness in \cite{BohnDurin2018SciRep} suggests that the potential
constant disorder line intersects the critical line at a point 
corresponding to a thickness 100nm, leaving $2\times$ and $5\times$ 
thinner samples on the left of the critical line. Since none of these
samples are infinitely thin, the corresponding line should be high
enough in the phase diagram, for example,
such that the theoretical critical thickness can be close to the
theoretical $l_{ref}\approx L/16$, as illustrated in Fig.\ S3 in SI.  
Note that these quantitative comparisons serve only as the RFIM description of the
actual intrinsic disorder in that sample, in view of the observed change of the
scaling exponents. Then observing that the reference thickness 
corresponds to 100nm in these samples, we can place all other
experimental data 
relative to
this point, see Fig.\ S3 in SI.
Hence, for the two thinner samples, the measured exponents should be
dominated by the second slopes $(\tau_2,\alpha_2)$; note that the
measured values are  in agreement with the
theoretical ones shown in the Table\ S-I for the loop-integrated
distributions. 
Then for the thicker samples, the exponents of the first slope
$(\tau_1,\alpha_1)$ and in the central hysteresis loop seem to
dominate the observed experimental distributions.
Note that in this region, the distance between the critical line and
the considered fixed disorder line is rather small; the corresponding
theoretical exponents are also highlighted in the Table\ S-I, for better
comparisons. The simulated avalanche distributions
along the fixed disorder line are also shown in Fig.\ S2a,b in SI; in this case, the
second slope, which is apparent in the critical avalanches, is
practically lost in the sub-critical disorder because of a large number of system-size
avalanches (resulting in the peak at the end of the distribution).
In the amorphous samples, however, the apparent disorder line seems to
be even higher, see Fig.\ S3 in SI. The exponents thus coincide with the ones $(\tau_2,\alpha_2)$ up
to  $2\times l_{ref}$, see Table\ S-I. Note also that the exponents in \cite{koreans_JApplPhys2008} 
measured for the very thin films of varied composition $x <0.5$,
including $x=0$,  are close to the second slopes
$(\tau_2,\alpha_2)$ estimated in the hysteresis loop centre, which are
listed in top-right part of the Table\ S-I.  More experimental
results shown in Fig.\ S3 in SI also confirm this systematic pattern of the
avalanche statistics.
Moreover, our results suggest that in such
thin samples at and around the critical disorder the multi-fractal
features of the BHN signal change with the sample thickness and that
they can depend on the time interval in which the scaling region is
considered.  At the critical disorder
line, some intervals have virtually mono-fractal behaviour, at the same time, the
surrounding intervals can show apparent multi-fractality.

We have demonstrated that a new type of collective
dynamics can arise on the hysteresis loop due to the interplay of the
sample geometry and critical fluctuations, studied along the
critical-disorder line for different thicknesses, interpolating
between the strictly two-dimensional and the three-dimensional systems. 
The geometry of the sample has a profound impact on the magnetic response of
sufficiently thin systems, and it is manifested in a time-scale dependent
multi-fractality of Barkhausen noise and double power-law distributions
of the magnetisation-reversal avalanches, both of which differ from
those known in the limiting cases of two-dimensional and
three-dimensional geometry. The main cause of these new critical properties can be associated with the pinning of the domain walls at
the open boundaries of thin samples, which thus constrain the
avalanche shape and its propagation by effectively changing the role of intrinsic disorder, and causes an intermittent appearance of
large and small avalanches even in the central segment of the
hysteresis loop. These effects are most apparent in the shape of
\textit{critical avalanches}, but they can also be observed in the range of disorders
close to the critical line.  These findings are in agreement with
some recent experimental results, in a restricted range of the parameters
where the comparison is permitted by given experimentally accessible conditions. In addition to a wide range of
samples with different sizes and thicknesses, the presented numerical
results include the exact two-dimensional samples and the whole range
of time scales, which are beyond the reach of  the laboratory experiments.
In this regard, our results can serve as a guide for further experimental investigations; they also reveal new features 
of the domain-wall stochasticity in thin ferromagnetic films, which
are important for developing new technological applications.

\acknowledgments
BT acknowledges the financial support from the Slovenian
Research Agency (research code fund- ing number P1-0044). SM, SJ, DjS
acknowledge the financial support from the Ministry of Education,
Science and Technological Development of the Republic of Serbia, 
Project No. 171027.

\end{document}